# Space-Time Distribution of G-band and Ca II H-line Intensity Oscillations in Hinode/SOT-FG Observations

**J. K. Lawrence[1] and A. C. Cadavid[1]**

[1]Department of Physics and Astronomy, California State University, Northridge,
18111 Nordhoff Street, Northridge, California 91330-8268, USA
(e-mail: *john.lawrence@csun.edu*, *ana.cadavid@csun.edu* )

**Abstract.** We study the space-time distributions of intensity fluctuations in 2 – 3 hour sequences of multi-spectral, high-resolution, high-cadence, broad-band filtergram images of the Sun made by the SOT-FG system aboard the Hinode spacecraft. In the frequency range $5.5 < f < 8.0$ mHz both G-band and Ca II H-line oscillations are suppressed in the presence of magnetic fields, but the suppression disappears for $f > 10$ mHz. By looking at G-band frequencies above 10 mHz we find that the oscillatory power, both at these frequencies and at lower frequencies too, lies in a mesh pattern with cell scale 2 – 3 Mm, clearly larger than normal granulation, and with correlation times on the order of hours. The mesh pattern lies in the dark lanes between stable cells found in time-integrated G-band intensity images. It also underlies part of the bright pattern in time-integrated H-line emission. This discovery may reflect dynamical constraints on the sizes of rising granular convection cells together with the turbulence created in strong intercellular downflows.

# 1. Introduction

Of fundamental interest in solar physics are the mechanisms by which non-thermal energy passes from the turbulent convection zone through the photosphere, to the chromosphere and beyond. Analysis of the physics of these processes around the photosphere-chromosphere interface requires data sequences at the finest possible spatial and temporal scales. Vertically, we focus on the margin between solar features seen in optical continuum images or photospheric band passes like the CH G-band or in Fe I filtergrams, on the one hand, and co-aligned Ca II H- or K-line filtergrams, on the other. Thus we make use of high-resolution, high-cadence time series of broadband filter image (BFI) filtergrams and magnetic images made in these wavelength bands. Modern analytical techniques like temporal and spatial wavelet transforms also are useful. With them we can identify, categorize, and perhaps understand, the physical processes that connect photospheric magnetism and dynamics to chromospheric structure, oscillations and heating.

In earlier work (Cadavid *et al.*, 2003; Lawrence *et al.*, 2003) we studied a nine hour data set taken 30 May 1998 with the Swedish Vacuum Solar Telescope (SVST) (Berger and Title, 2001). The data included a series of 1541 near simultaneous G-band and Ca II K-line filtergrams with cadence 21 s and pixel scale averaged to 0.24 Mm. Also available were magnetograms of lower cadence and resolution. Among other results in these papers, we identified two categories of mutual oscillation or fluctuation of the G-band and K-line intensities (Lawrence *et al.*, 2003). The first involved acoustic oscillations with equal G-band and K-line frequencies both above and below the acoustic cutoff frequency around 5 mHz. A less straightforward category featured mutually exclusive frequency ranges of G-band and K-line fluctuations ($f_G > 5$ mHz, and $f_K < 4$ mHz).

The Solar Optical Telescope (SOT) (Tsuneta *et al.*, 2007) and its focal plane package (Tarbell, 2006) on the Hinode spacecraft (Kosugi *et al.*, 2007) can take multispectral, broad-band filtergram images, including the G-band and Ca II H-line, at high resolution and at rapid cadences that allow the study of frequencies well above 10 mHz. The sequence lengths are of the order of 2 – 3 hours, and there are several data sets with fields of view up to 80 Mm. Pixel scales are generally 0.079 Mm. There is no need for numerical seeing corrections of spacecraft images. Simultaneous magnetic (*V/I*) images also are available in many cases, although these require spatial remapping and co-registration with the filtergrams.

Using the improved spatial resolution and stability of these data, we have been able to advance the results of our previous work as described more fully below. Notably, we find that for $f > 5.5$ mHz the G-band spectral power tends to appear in a cellular or web like pattern with characteristic spatial scale 2 – 3 Mm and correlation times of the order of hours. This same pattern is seen in the H-line spectral power maps at lower frequencies $2.6 < f < 4.0$ mHz. The web pattern appears in the lanes between stable features in time-integrated G-band intensity images, and they underlie some parts of the integrated H-line intensity. At the high frequencies where we see the cellular pattern the overall spectral power is very low, but the pattern is strong also at lower frequencies, although harder to see.

In Section 2 we introduce the data sets used in the present work and indicate how the data were prepared. Section 3 describes the data analysis procedures and presents the results thereof. Results are discussed and conclusions are drawn in Section 4.

## 2. Data and Data Processing

The work presented in this paper is based on several high-resolution, high cadence time series of G-band, Ca II H-line and Magnetic Hinode/SOT-FG images of the Sun (Ichimoto *et al.*, 2008; Shimizu *et al.*, 2008; Suematsu *et al.*, 2008; Tsuneta *et al.*, 2008). The data were downloaded from the DARTS website and were processed to Level 1 with the SSW FG-PREP package in IDL.

**2.1 Data**

For purposes of presentation, we mainly focus on BFI data taken on 14 April 2007 from UT 14:04:30 to UT 16:54:58. For the G-band (430.50 nm, band pass 0.8 nm) and Ca II H-line (396.85 nm, band pass 0.3 nm) FG images, the field of view was $40 \times 40$ Mm, and the images were $2 \times 2$ square-averaged to $512 \times 512$, for a pixel scale of 0.079 Mm. The heliocentric cosine of the image center was $\mu = 0.80$. The magnetic data were Shuttered FG *I* and *V* Fe I 630.2 nm, band pass 0.6 nm, *V/I* images with field of view $59 \times 59$ Mm in a $2 \times 2$ binned format of $512 \times 512$ pixels. The magnetic images are not calibrated to magnetic units but are in units of normalized *V* profile asymmetry. The magnetic images were dilated and translated by us and thereby co-registered with the G-band and H-line filtergrams.The cadence of the images of each type was 21 s. In each camera cycle the H-line images were taken 3.2 s after the G-band, and then the magnetic data were taken 1.7 s after the H-line or 4.9 s after the G-band.

The data are *x-y-t* cubes. Figure 1 gives sample *x-y* images of these data at the start of the time series. Figure 2 gives *x-t* slices of the data cubes at $y = 22$ Mm.

A similar sequence to the above was taken on 30 March 2007 from UT 06:14:37 to UT 09:24:24. The field of view was $80 \times 40$ Mm in $1024 \times 512$ pixel format and with pixel scale 0.079 Mm. The heliocentric cosine $\mu = 0.83$. The image cadence was 35 s with the H-line images following the G-band by 3.2 s and the magnetic images following the H-line by 3.3 s or 6.5 s after the G-band.

A third sequence was taken on 9 April 2007 from UT 15:00:00 to UT 16:59:20. The field of view was $80 \times 40$ Mm in $1024 \times 512$ pixel format and pixel scale 0.079 Mm. The images were at disk center with $\mu = 1.00$. The image cadence was 30 s with the H-line images following the G-band by 3.2 s and the magnetic images following the H-line by 1.7 s (or 4.9 s after the G-band).

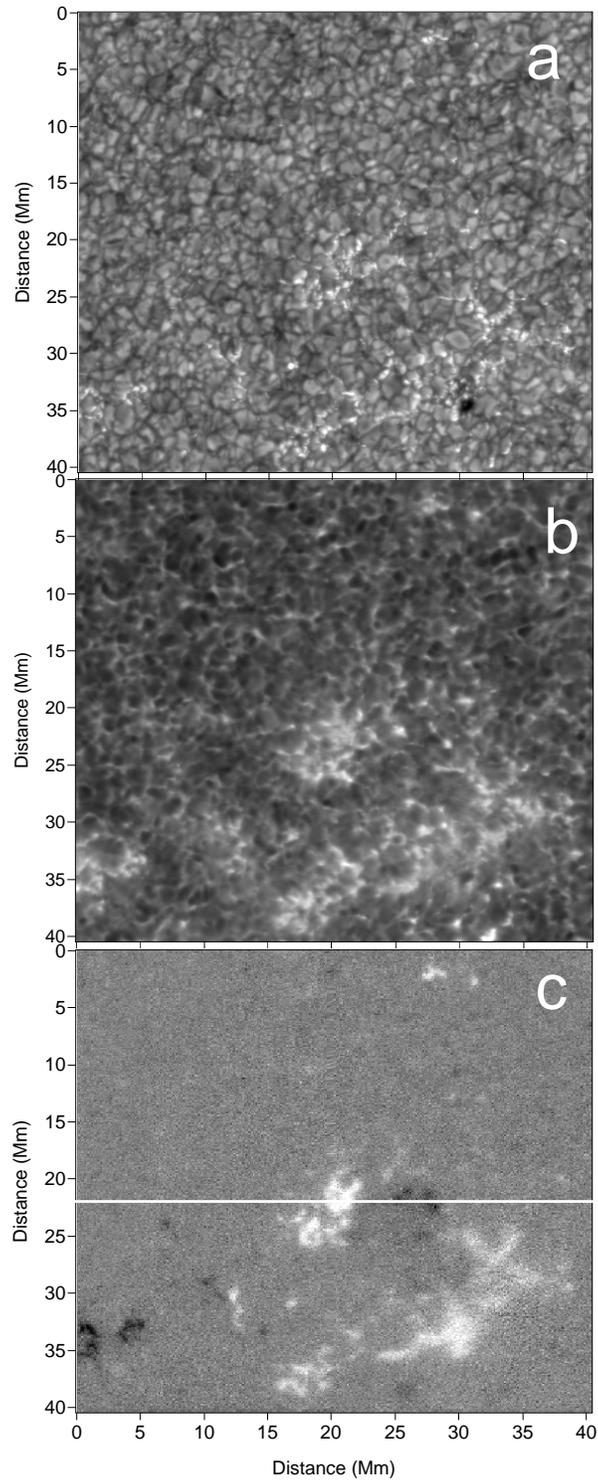

**Figure 1.** Sample images taken 14 April 2007 at UT 14:04:30. (a) Sample G-band FG image. (b) Sample Ca II H-line FG image. (c) Sample magnetic (shuttered *V/I*) image. The magnetic images were dilated and translated to match the FG images in Figures 1a,b. The white line at $y = 22$ Mm marks the location of time slices shown in Figures 2 and 6.

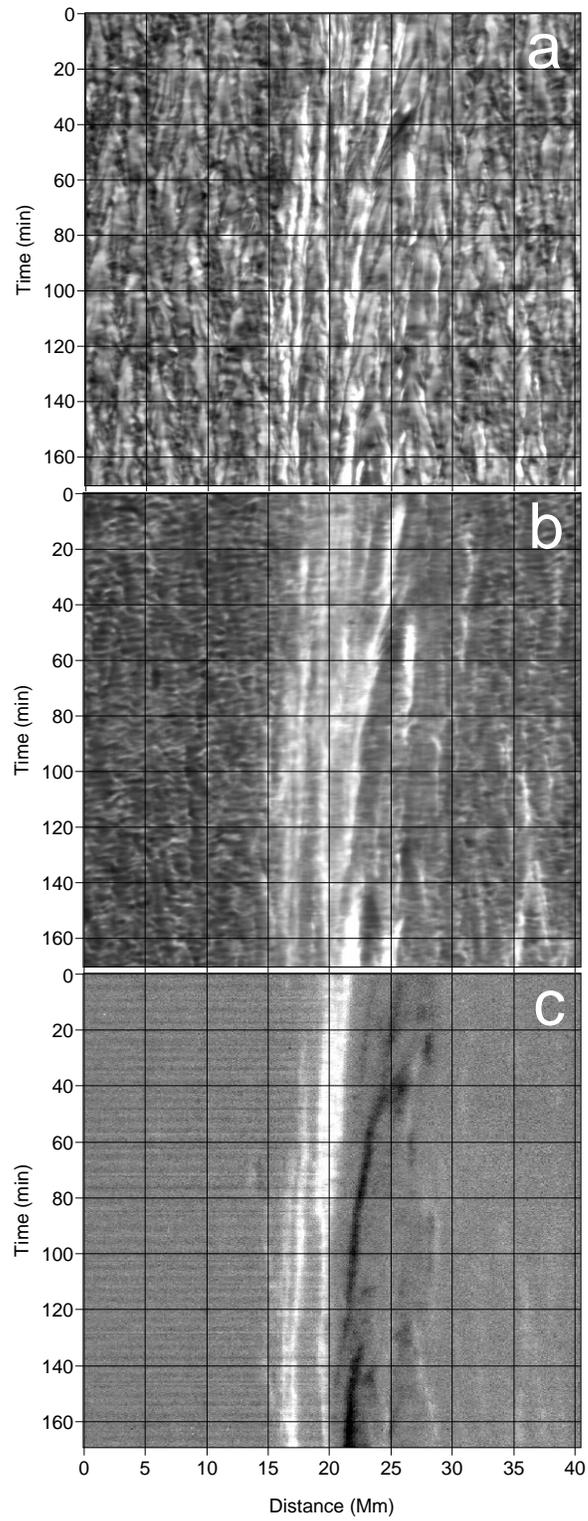

**Figure 2.** *x-t* slices at *y* = 22 Mm of (a) G-band intensity data cube, (b) H-line intensity data cube, and (c) the magnetic (*V/I*) data cube. The time, in minutes, increases downward.

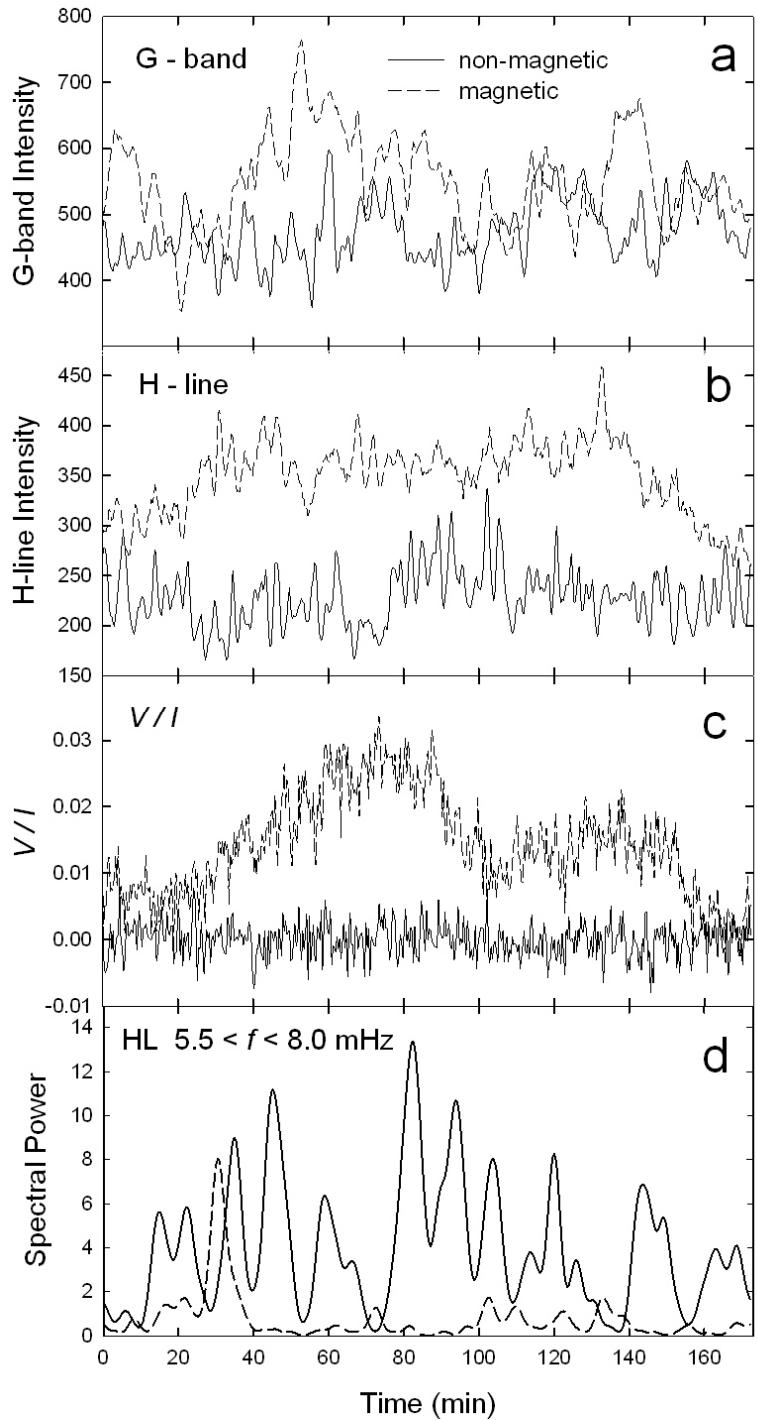

**Figure 3.** Time traces of (a) G-band intensity, (b) H-line intensity, (c) *V/I* (normalized *V* asymmetry), and (d) H-line Morlet wavelet power in the 5.5 – 8.0 mHz frequency band (*T* = 2 – 3 min) at two spatial locations vs time in minutes. The spatial locations are at *y* = 22 Mm corresponding to the time slices of Figure 2. The nonmagnetic location (solid lines) is at *x* = 10 Mm, and the magnetic location is at *x* = 19 Mm. Signal units are those of the SOT–FG data.

### 2.2 Wavelet Spectra

Our principal focus is on intensity oscillations in the G-band and H-line images. We therefore generated time series of oscillatory power images from the corresponding intensity images. To capture both spectral and time information we used Morlet wavelet transforms calculated with an IDL package due to Torrence and Compo (1998). This is available at the website http://paos.colorado.edu/research/wavelets. The Morlet wavelet mother function is a plane wave modulated by a Gaussian envelope: $\psi(t) = \pi^{-1/4} \exp(i\omega t) \exp(-t^2/2)$.

The time integrated wavelet spectrum is equal to the Fourier spectrum smoothed by the Fourier transform of the wavelet mother function at each scale. In Figure 4 we present *x-y* images of time-integrated G-band wavelet power. To make contact with the work of Vecchio *et al.* (2007) and their analysis of Doppler oscillations in the Interferometric BIdimensional Spectrometer (IBIS) data we have further integrated the wavelet transformed images over spectral bands with (a) frequency $f < 1.2$ mHz (or period $T > 13.8$ min), (b) $2.6 < f < 4.0$ mHz ($4.2 < T < 6.4$ min), and (c) $5.5 < f < 8.0$ mHz ($2 < T < 3$ min). This last band was named the "high frequency" (HF) band by Vecchio *et al.* (2007). The 21 s cadence of our Hinode data set allows us also to define a "very high frequency" (VHF) band including frequencies $10 < f < 24$ mHz ($42 < T < 100$ s). Figure 5 shows the corresponding H-line spectral power images.

The use of wavelet transforms in time allows us, as in the case of the intensity images, to examine the temporal behavior of the oscillation data. This is shown as *x-t* slices in Figure 6 and as single magnetic and non-magnetic time traces in Figure 3d.

## 3. Results

### 3.1 Magnetic Suppression of Oscillatory Power

In the H-line time traces in Figure 3b we see that, although the overall H-line emission is enhanced when magnetic field is present, the amplitude of the short-term fluctuations is reduced. The same is true to a lesser extent in the G-band in Figure 3a. Figure 3d shows the Morlet wavelet power of the two H-line time traces in the 5.5 – 8 mHz (2 – 3 minute) band. In the non-magnetic case there is a series of about a dozen ˜ 10 minute bursts of wavelet power. In the magnetic case there is only one such burst, and that is at a time when the magnetic field is not very strong.

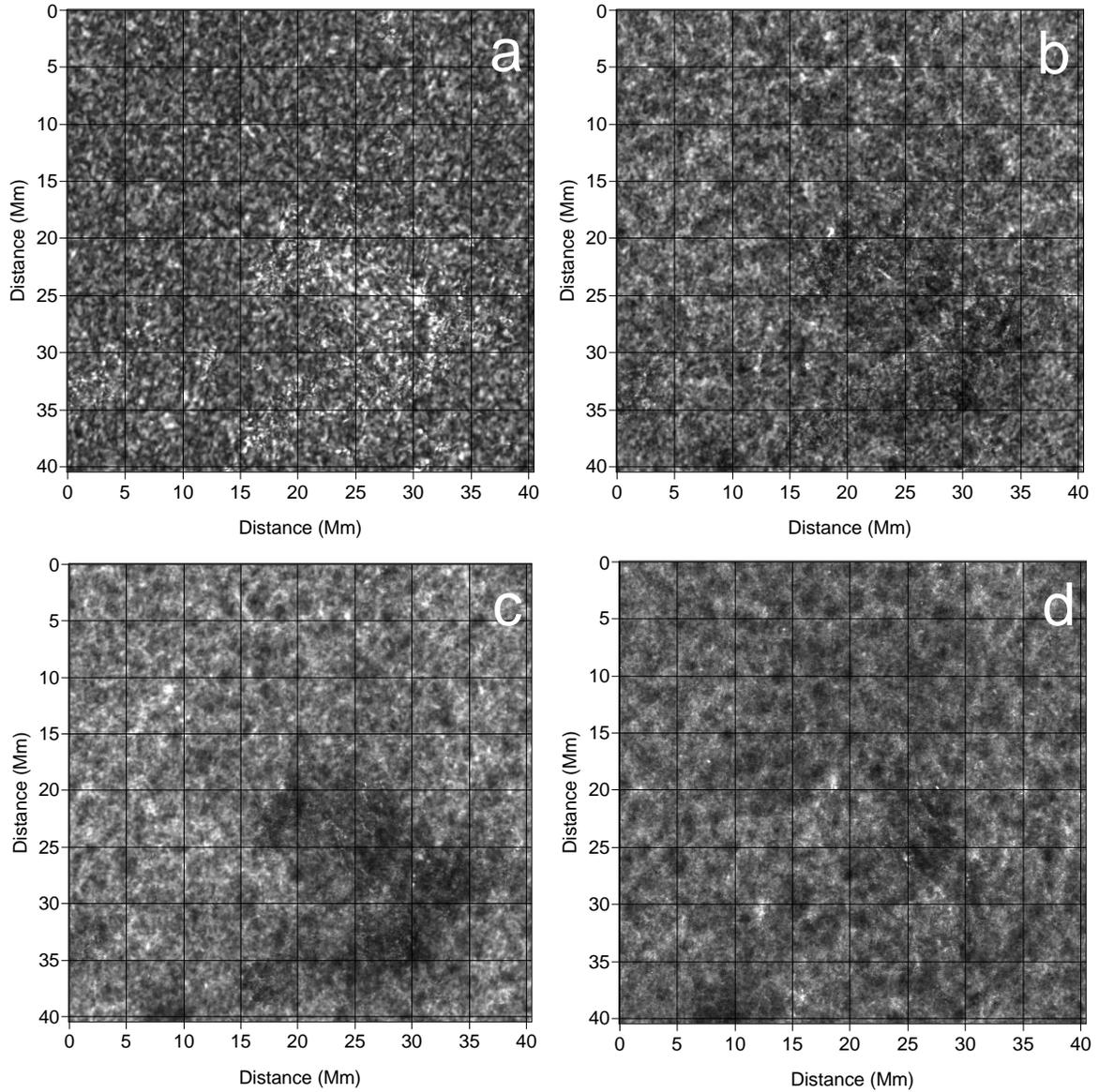

**Figure 4.** Images of time integrated G-band spectral power for the data set in Figures 1 and 2. The spectral bands are (a) frequency $f < 1.2$ mHz or period $T > 13.8$ min, (b) $2.6 < f < 4.2$ mHz ($4.2 < T < 6.4$ min), (c) the HF band $5.5 < f < 8.0$ mHz ($2 < T < 3$ min), and (d) the VHF band $10 < f < 24$ mHz ($42 < T < 100$s).

By comparing the H-line oscillatory power map in Figure 5c to the magnetic map in Figure 1c we see that the HF $5.5 – 8.0$ mHz ($2 – 3$ min) H-line oscillatory power is strongly reduced in the presence of magnetic fields. And in Figure 4c there is suppression in the HF

G-band oscillatory power as well. There also is reduction in both G-band and H-line oscillatory power through time in Figures 6a, b as seen in comparison to the magnetic *x-t* time section in Figure 2c.

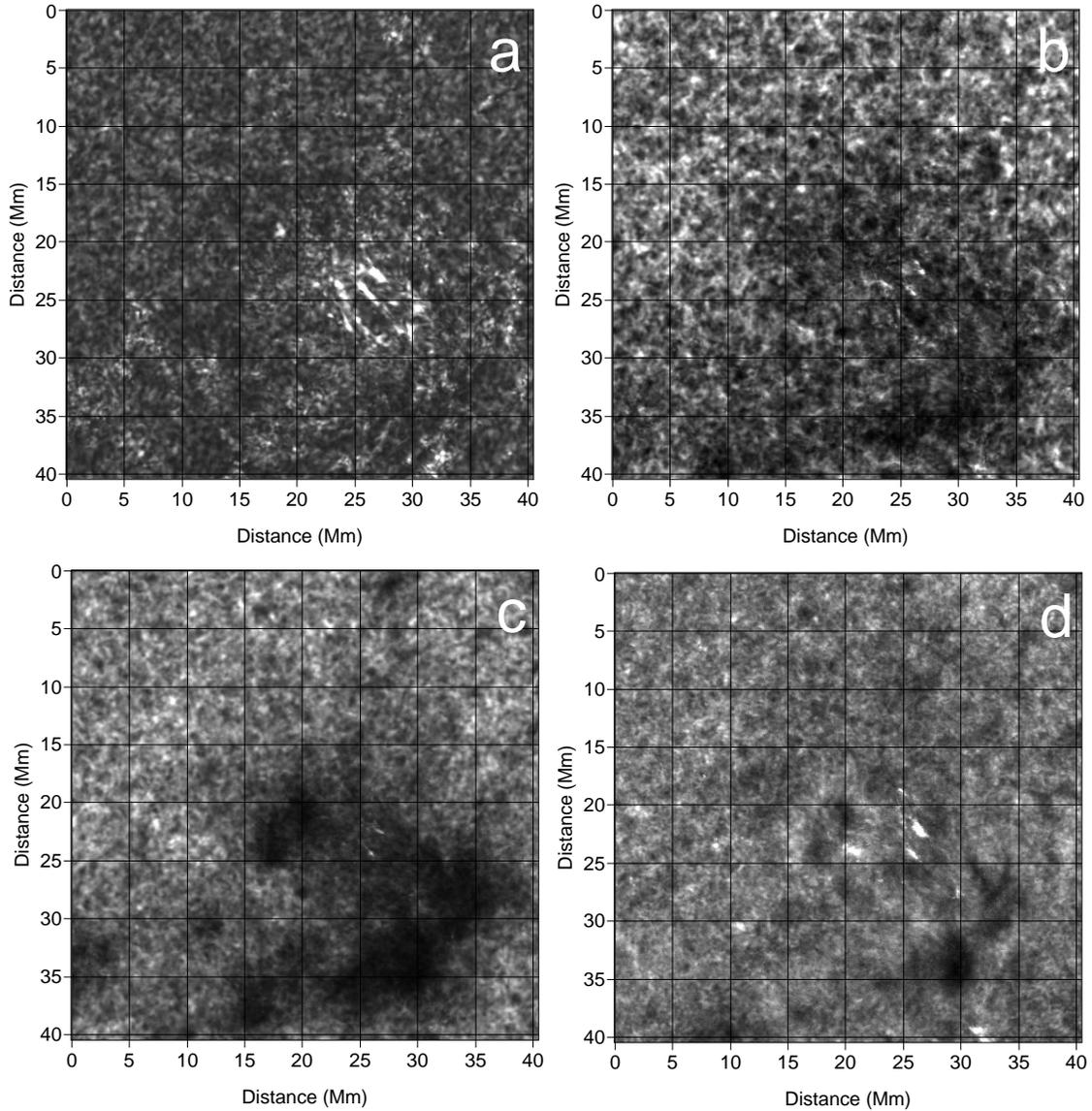

**Figure 5.** Images of time integrated H-line spectral power for the data set in Figures 1 and 2. The spectral bands are (a) frequency $f < 1.2$ mHz or period $T > 13.8$ min, (b) $2.6 < f < 4.2$ mHz ($4.2 < T < 6.4$ min), (c) the HF band $5.5 < f < 8.0$ mHz ($2 < T < 3$ min), and (d) the VHF band $10 < f < 24$ mHz ($42 < T < 100$s).

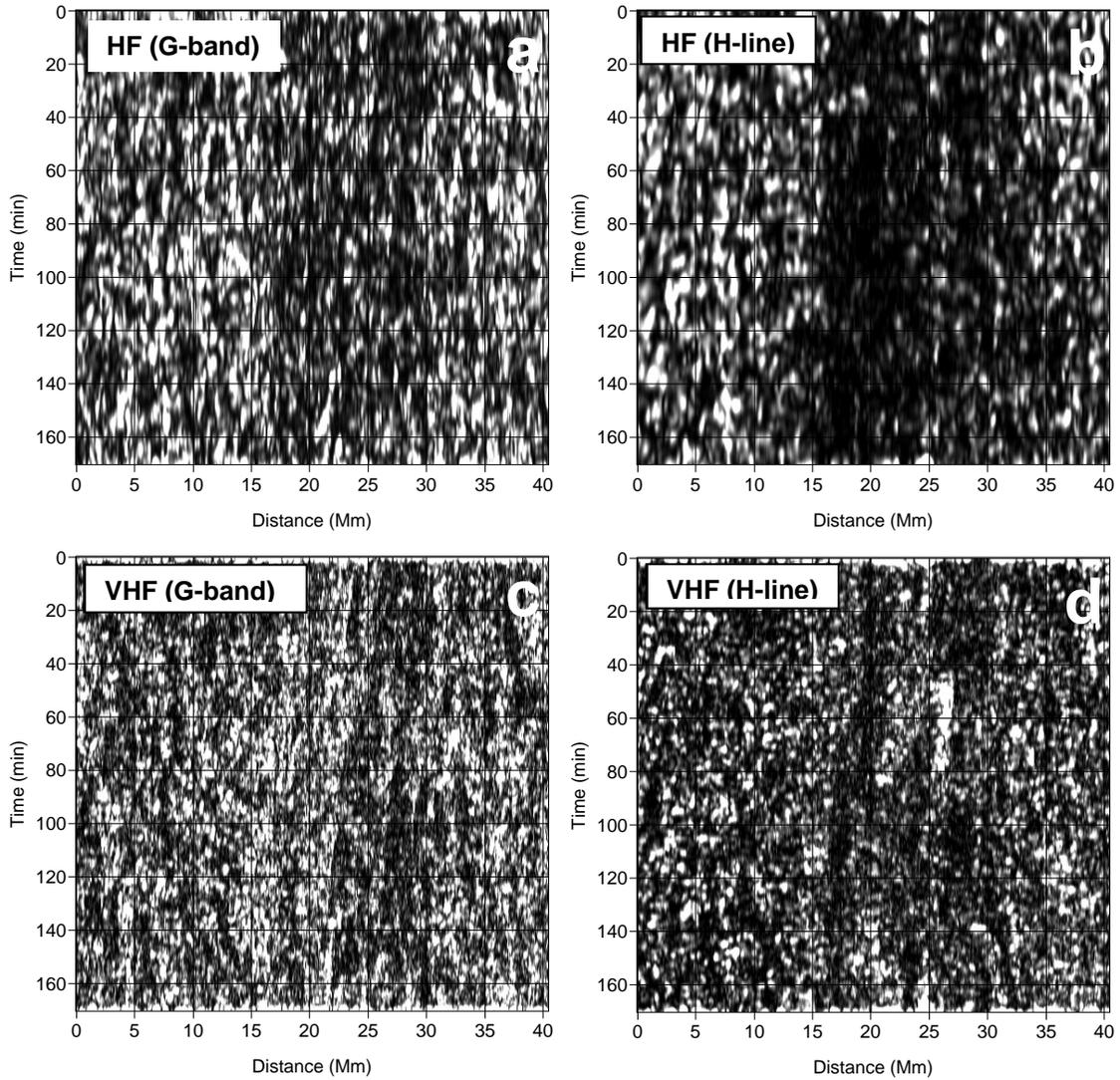

**Figure 6.** *x-t* images of wavelet power data cubes. Here the horizontal axes show distance in megameters, and the vertical axes show time, increasing downward, in minutes. (a) HF ($2 < T < 3$ min) in the G-band, (b) HF in the H-line, (c) VHF ($42 < T < 100$s) in the G-band, and (d) VHF in the H-line. Compare these to Figure 2c showing the corresponding magnetic fields.

Interestingly, the inhibition of oscillatory power is not present in the VHF images with $10 < f < 24$ mHz ($42 < T < 100$ s) seen in Figures 4d and 5d. In the corresponding *x-t* sections of Figures 6c, d some weak suppression of the oscillations still is visible, though its location changes in time which disguises its presence in the time-averaged images in Figures 4 and 5. The presence of structure is easier to see in the *x-t* projection.

## 3.2 VHF Cellular Pattern at Scale 2 – 3 Mm

In Figure 4d the VHF G-band oscillatory power outlines a pattern of roughly circular cells with diameters 2 – 3 Mm. This pattern also can be made out in Figure 4c showing the HF oscillatory power.

The cells we find are composed of relatively dark centers surrounded by lighter network. Conversely, a field of granules is made up of light centers in a dark intergranular network. Both kinds of data are open to studies of their spatial scales by means of "Mexican Hat" (MH) wavelet spectra. As the name implies, the wavelet mother function comprises a central peak surrounded by a negative ring. A standard form for this is the function: $MH(2) = -\nabla^2 \Phi(r)$, where $\Phi = \exp(-r^2/2s^2)$ is a circularly symmetric Gaussian of width $s$. Still better scale resolution is achieved by means of the sixth-order Mexican Hat: $MH(6) = -\nabla^6 F(r)$ (Lawrence, Cadavid and Ruzmaikin, 1999, 2001). Normalized G-band spectra are shown in Figure 7 and for H-line in Figure 8. These spectra respond both to bright centers surrounded by dark rings and dark centers surrounded by bright rings. The two sub-spectra, when separated, resemble one another rather closely. The spectra in Figures 7 and 8 are normalized to a sum of unity. Figure 9 illustrates the actual relative power in the four frequency bands.

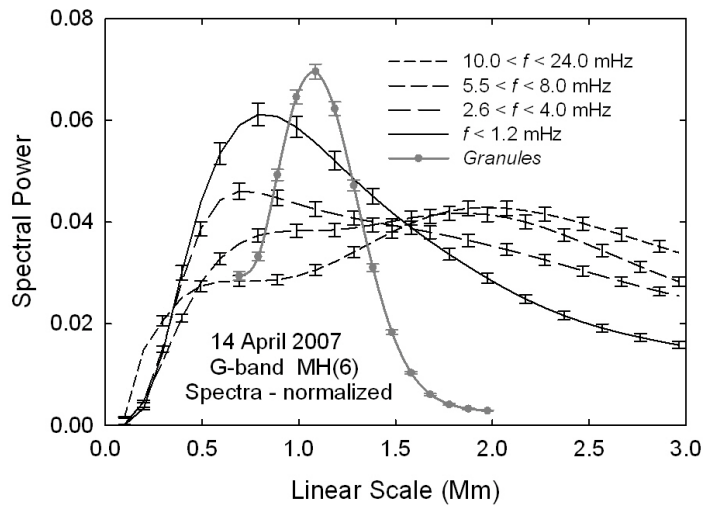

**Figure 7** Normalized, MH(6) spatial power spectra of Figures 4a – d indicating the spatial scales of G-band oscillatory power in the four frequency bands and from a field of granules in G-band intensity.

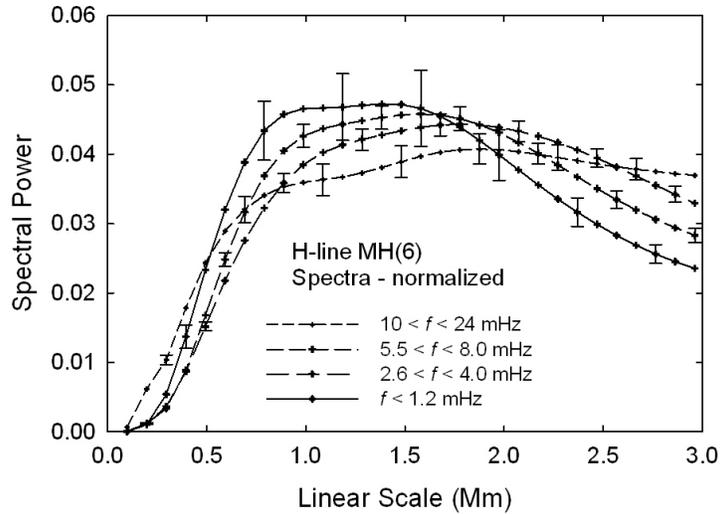

**Figure 8** Normalized, MH(6) spatial power spectra of Figures 5a – d indicating scales of H-line oscillatory power in the four frequency bands.

The error bars in Figures 7 and 8 and Figure 13 below are based on the variance of spectral power values in each linear scale bin and an estimated spatial correlation length of 10 pixels. They are checked by comparing spectra from different image areas.

The non-normalized wavelet power in Figure 9 is plotted on a logarithmic scale, and the total VHF power is thus of the order of 1% or less than that in the other bands. Therefore, although the 2 – 3 Mm cells are easiest to see in the VHF spectral image, most of the associated spectral power resides in the lower frequency bands. The 2 – 3 Mm scale also appears in the G-band HF spectrum and in the H-line in HF and also for $2.6 < f < 4.0$ mHz.

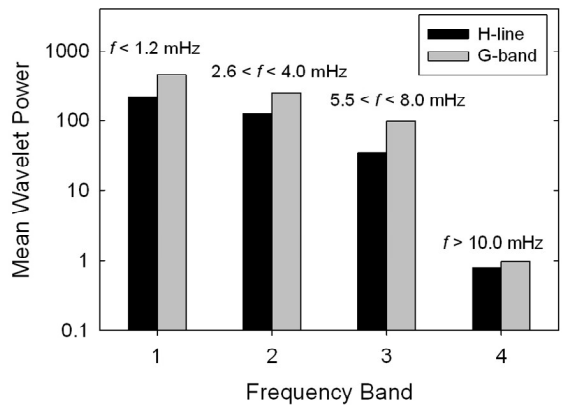

**Figure 9** Integrated spectral power in the four G-band and H-line oscillatory frequency bands. The VHF power is 1% or less than that in the other bands.

The key result from the analysis of this Section is that we find a cellular network, that is outlined visually in the VHF G-band oscillatory power, with scale $2 < D < 3$ Mm.

This is distinctly larger than the typical granular scale peaking at 1.2 Mm as indicated in Figure 7. This scale also is distinctly smaller than that of mesogranules as seen in horizontal flows (Rieutord *et al.*, 2000). UeNo and Kitai (1997) found 5 minute oscillatory power to be concentrated in the downflows surrounding mesogranules of size 15 arcsec or about 12 Mm. The structures we see here are smaller than that by about a factor of 5.

**3.3 Correlation Time of Cell Patterns**

The images in Figures 4c, d and in Figures 5c, d are sums of VHF wavelet power over nearly 3 hours time. Thus the cell patterns may be correlated over at least a similar time. Figures 6c, d show that the wavelet power is intermittent, but they also show signs of spatial patterns that repeat in time. This points to the possibility of patterns of wavelet power that come and go but are correlated over time.

To test this idea, we took a series of the VHF wavelet images spaced 3 minutes apart. Each individual image was standardized to zero mean and unit variance. The correlation of one Image *A* with another Image *B* is $C_{AB}(t_1,t_2) = \sum_{ij} I_A(i,j,t_1) \cdot I_B(i,j,t_2)$. By correlating all the images in the stack with all the others, we can build up a correlation versus time plot. In Figure 10 we show plots for the 30 March, 9 April, and 14 April 2007 data sets versus time interval. The error bars represent the scatter in binned correlations divided by the square root of the number of entries in each bin. The correlation drops from 100% at $\Delta t = 0$ to 20-30% in about 3 min and then to about 1% in 40 min. The 1% correlation remains significantly non-zero to about time 120 min.

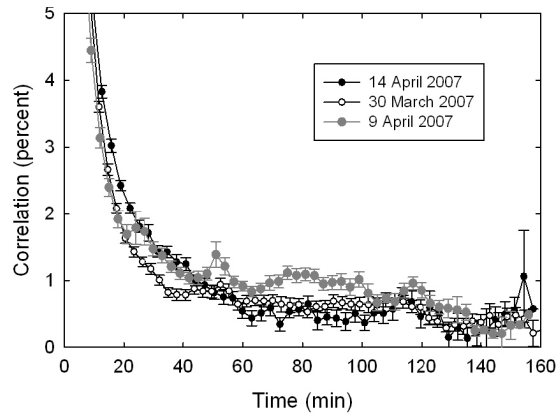

**Figure 10.** Mean image correlations, in percent, of VHF G-band spectral images versus time interval between the images in minutes.

## 3.4 Principal Components Analysis of a Space-Time Data Cube

Another approach to this question is offered by principal component analysis (PCA). Here, the first "empirical orthogonal function" (EOF) is that linear combination of the images that is most strongly correlated to the data as a whole. We find this by diagonalizing the correlation matrix described in section 3.3.

Figure 11, which presents the spectrum of eigenvalues, indicates that the leading mode stands out strongly from the trend of the higher eigenvalues, and that it accounts for 4% of the total data variance.

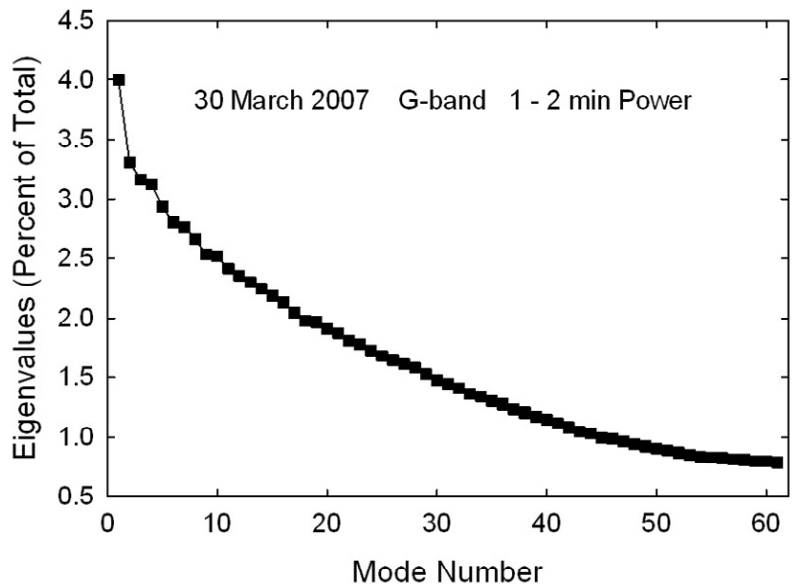

**Figure 11.** Eigenvalues of the correlation matrix of the set of 61 G-band VHF spectral power images taken 30 March 2007. These are in units of the percent of the total image variance captured by each mode.

Figure 12a shows a sample single G-band intensity image from the 30 March 2007 data set, and Figure 12b displays the EOF of the first mode of the corresponding VHF power. The first EOF shown in Figure 12b clearly displays the same cellular pattern we saw previously in the time-averaged VHF G-band spectral image of Figure 4d. In fact, Figure 12b is effectively identical to the time-averaged VHF G-band spectrum for the March 30 data set. Figure 13 gives MH(6) spatial power spectra for Figure 12a (dashed) for a G-band intensity image of a field of granules, and for Figure 12b (solid) for the first EOF of images of G-band VHF oscillatory power. The dominant scales in the two images are clearly resolved.

The dominant scale in the VHF image is near 2.0 Mm with a clear shoulder at smaller scale around 0.5 Mm, while the dominant scale of the granular image is 1.2 Mm.

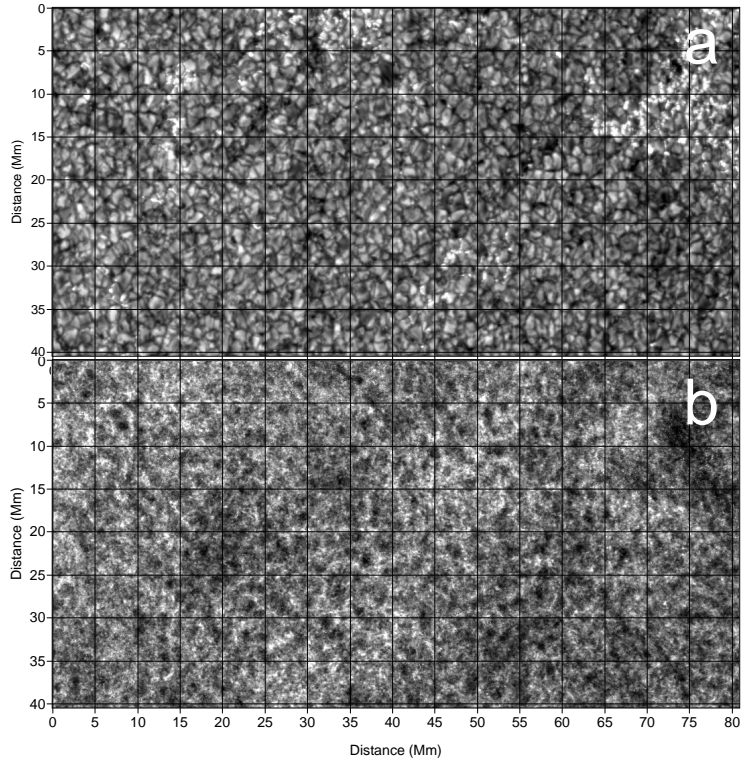

**Figure 12.** (a) Sample G-band intensity image from the 30 March 2007 data set. (b) the $1^{st}$ EOF of the PCA analysis of the G-band VHF spectral power data. Spatial units are megameters.

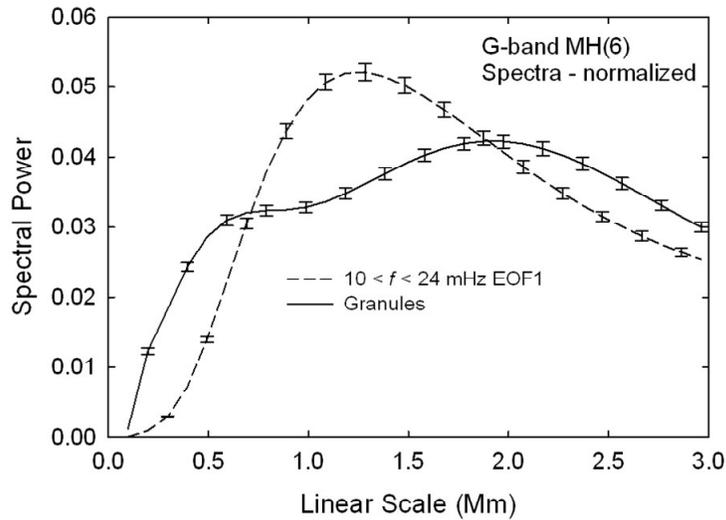

**Figure 13.** MH(6) spatial power spectra for Figure 12a (dashed) and Figure 12b (solid) versus linear scale in megameters.

Figure 14 shows the first principal components (PCs) for the 14 April and the 30 March 2007 data sets. These indicate the levels of correlation in percent of the individual G-band VHF images with the first EOFs which, in turn, represent the cellular patterns of G-band VHF spectral power. In the case of the 9 April 2007 data, the cellular pattern is divided between the first two EOFs so its PC is not included here.

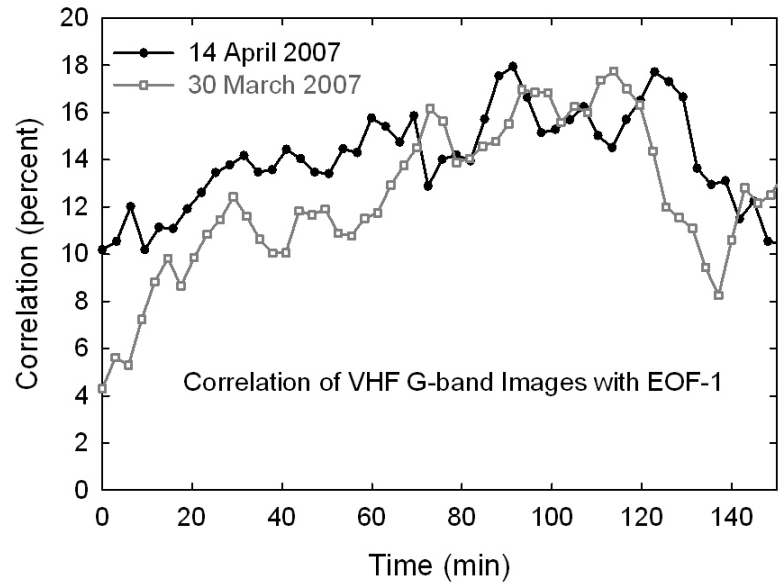

**Figure 14.** Plot of the first PCs for the 2007 March 30 and 2007 April 14 VHF G-band data sets. These indicate the projection of the standardized first EOFs onto the individual standardized VHF G-band images through time. These give the correlations of the individual images with the first EOFs which represent the observed cellular patterns in G-band VHF spectral power.

**3.5 Connection of Spectral Power to Other Data**

As an aid to interpretation of the physical origin of the cellular pattern we compare the VHF (and HF) G-band spectral power images to other data, in particular to the time-averaged G-band and H-line intensity images.

By visual inspection, one can see anticorrelation between Figures 15a, b: the brighter areas of spectral power tend to come in the dark gaps between the time-averaged, G-band granular intensity features. For example, the light feature in Figure 15a at $x = 3$ Mm and y = 5 Mm is dark in Figure 15b. Comparing Figures 15b, c shows some partial positive correlation between the brighter areas of VHF power and H-line emission. The features at $x = 6$ Mm, $y = 5$ Mm and at $x = 9$ Mm, $y = 4$ Mm are bright in Figures 15b, c while the light feature in Figure 15c at $x = 3$ Mm, $y = 5$ Mm is dark in Figure 15b.

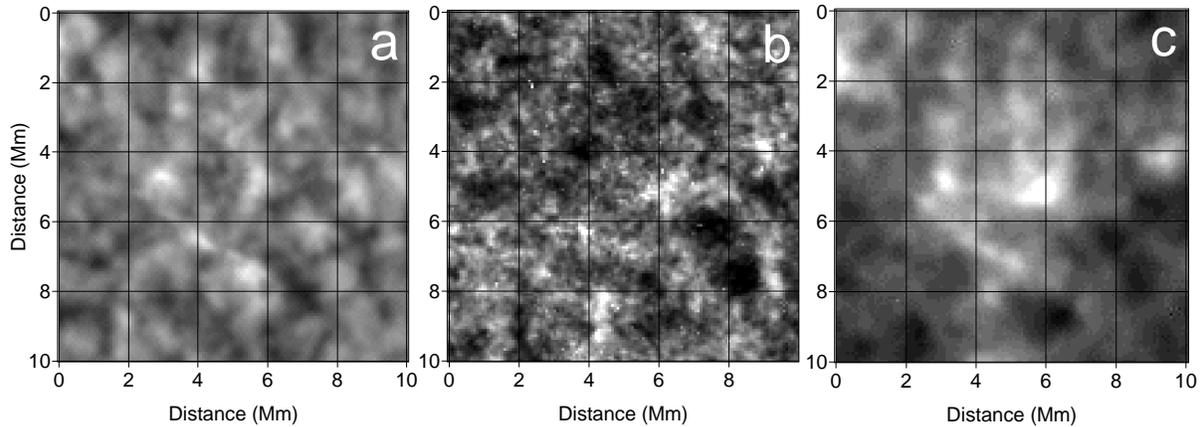

**Figure 15.** Image details, each 10 Mm on a side, taken from the 30 March 2007 data set of (a) time-integrated G-band intensity, (b) G-band VHF spectral power, and (c) time-integrated H-line intensity.

A more quantitative, but less detailed, study of this is offered by plotting the G-band VHF spectral power in bins of integrated G-band and integrated H-line intensities. Figure 16 shows these plots for the 30 March and the 14 April 2007 data sets. Since the intensity data are in arbitrary units, we have arbitrarily binned them in equal steps from 0 to 100. That is, the range of observed intensity values, whatever it may be, is divided into 100 equal intervals that define 100 bins. Then the lowest intensities are placed in bin 1, the highest intensities in bin 100 and so on in between. These are then averaged within each bin. The last 20 bins, with relatively few entries, have been dropped from the plots. The bins with the greatest occupancy are those with the smallest standard errors. Note that in each Figure the bin occupancy is different between the G-band and the H-line plots. Because the G-band intensities are arbitrary, so are the VHF G-band power, and these have been adjusted to match one another. The widths of the 1s error bars give a general idea of where the main mass of the pixels contribute.

We can see in these plots that out to about 40 units, through the main mass of data pixels, the VHF G-band spectral power and the integrated G-band intensity are anti-correlated in both data sets. This range includes half of the image pixels. Conversely, out to about 20 units we see that the G-band VHF spectral power in positively correlated with the integrated H-line intensity for the less intense areas of H-line emission. This also includes half of the image pixels.

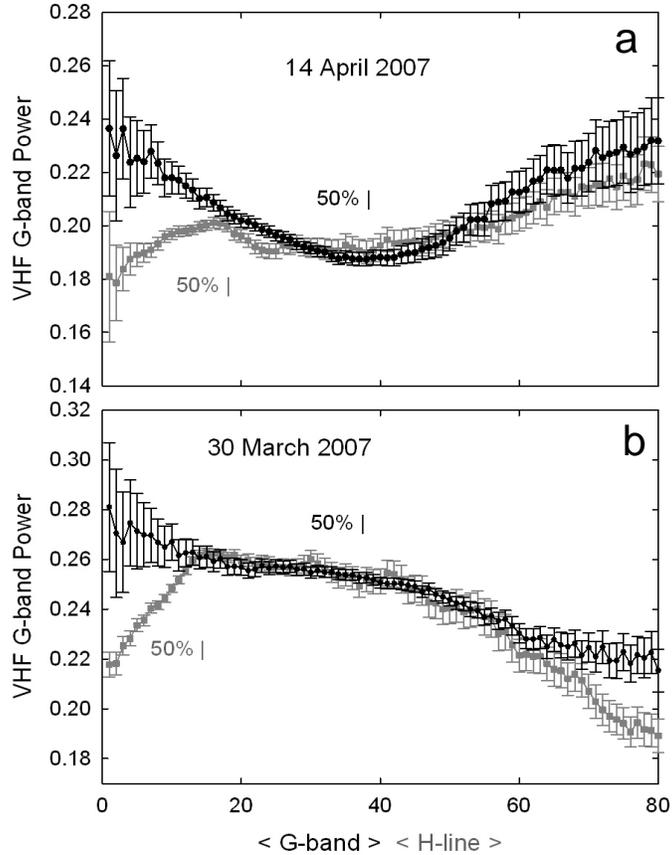

**Figure 16.** Plots of VHF G-band spectral power in bins of integrated G-band intensity (black) and integrated H-line intensity (gray). Plot (a) includes 30 March 2007 data, and plot (b) the 14 April 2007 data. The error bars are the 1s standard error of the contents of the bins. The vertical marks indicate the points where one-half of the image pixels fall on either side.

## 4. Discussion and Conclusions

**4.1 Magnetic Suppression of Oscillatory Power**

Inhibition of Doppler oscillatory power near magnetic fields, "magnetic shadowing," (Judge, Tarbell, and Wilhelm, 2000; see also Vecchio *et al.*, 2007) has often been related to mode conversion of upwardly propagating acoustic waves by the magnetic canopy. This seems perhaps unlikely here since the G-band is formed deep in the photosphere (Rutten, 1999) at 140 km above $t_{500} = 1$ (Steiner, Hauschildt, and Bruls, 2001) or as low as 60 km (Nordlund, Stein, and Asplund, 2009). This should lie below the magnetic canopy. Even much of the H-line emission is in the upper photosphere, perhaps 200 km above the G-band (Mitra-Kraev, Kosovichev, and Sekii, 2008). Thus the reduction of the HF intensity oscillations may indicate direct magnetic damping of the oscillations. We note that reduced convective motion in the presence of magnetic fields is commonly observed (*e.g.*, Title *et al.*, 1989)

## 4.2 Correlation Times of Cell Patterns

Some discussion has taken place in the literature about whether granular patterns do (Getling and Brandt, 2002; Brandt and Getling, 2008, Berrilli *et al.*, 2005) or do not (Rast, 2002) show long-term (hours) correlations. Our Figure 10 shows that VHF G-band images spaced 3 minutes apart in three different data sets maintain a positive correlation of about 1% for about 2 hours time. PCA permitted the calculation of the leading EOFs and corresponding PCs in the data sets. These leading EOFs are effectively identical to the VHF G-band cellular images. Figure 14 shows the first PCs which indicate that the correlations of the leading EOFs with the individual images are in the 10% to 20% range throughout the 2.5 hour time span covered. Our results therefore seem to support the position of Getling and Brandt (2002) and Berrilli *et al.*, (2005)

## 4.3 Connection of VHF Spectral Power to Other Data

Figures 16a, b show that for the lower half (in terms of the number of image pixels) of the integrated G-band intensity the VHF G-band oscillatory power is anticorrelated with the integrated G-band intensity. These plots are shown in black. On the other hand, the gray plots show that for the lower half of the integrated H-line intensity, the VHF G-band oscillatory power is positively correlated with the integrated H-line intensity. For the upper halves of the plots, in terms of pixel numbers, the integrated G-band and H-line intensities for the 30 March 2007 case both turn downward at high intensity while those for the 14 April case turn upward. The 14 April data set is more strongly magnetic than is the 30 March data set, so these high intensity pixels may show magnetic effects.

Comparison of Figures 16 a, b, which cover a small portion of the 30 March 2007 data 10 Mm on a side, shows that the oscillations tend to be located in the pattern of lanes between the stable, time-integrated G-band granular cells. At the same time, the locations of the VHF oscillations coincide, in part, with locations of time integrated H-line intensity and may therefore constitute one energy source for chromospheric emission.

## 4.4 Possible origin of the VHF Cellular Pattern at Scale 2 – 3 Mm

The MH(6) wavelet spectra in Figures 7 and 13 show that the 2 – 3 Mm size scales (diameters) of the cells outlined in the VHF G-band are smaller than and are clearly resolved from the circa 1.2 Mm scales found in a field of granules in G-band intensity. UeNo and Kitai (1997) found 5 minute oscillatory power to be concentrated in the downflows surrounding mesogranules of size 15 arcsec or about 12 Mm. The structures we see here are smaller than that by about a factor of 5.

Earlier studies have reported observations similar to ours. For example, Lawrence, Cadavid and Ruzmaikin (2001) studied a 2-hour sequence of MDI high-resolution Doppler images in quiet Sun at disk center by means of MH(6) wavelets and found, in addition to a granular field, the presence of convective cells of spatial scale 2 Mm and lifetimes longer than 45 min. They associated these cells with mesogranulation, although they are smaller than the typical 5 – 12 Mm scale attributed to mesogranules (see a review by Rieutord *et al.*, 2000). Vecchio *et al.* (2005) carried out a PCA of similar MDI Doppler data. They found that, after a mode representing the solar rotation, the next several important modes represented structures with spatial scale about 3 Mm and with long temporal scales from 2 to 14 hours. In a simulation of turbulent magnetoconvection, Cattaneo, Lenz, and Weiss (2001) found a similar network of "macrocells" with extended lifetimes.

Although the granular and macrocell scales are distinct, both scales lie within the full range of diameters found for granules in Figure 13. In an interesting study based on two-dimensional simulations, Ploner *et al.* (1998) classified granules into two groups based on their modes of disappearance: "dissolving" or "fragmenting." These authors found that the mean lifetime of fragmenting granules (7.7 min) was slightly shorter than that of dissolving granules (8.6 min). On the other hand, they found that the typical diameters were quite different: 0.9 Mm for the dissolving and 2.2 Mm for the fragmenting. These give a weighted mean diameter of 1.5 Mm. If we make some allowances for the differences between a 2D simulation and the actual Sun as processed by MH(6) wavelets, we suggest identifying the "fragmenting" granules with our 2 – 3 Mm VHF G-band cells and adopting a mean granular scale of 1.2 Mm.

Our observations differ from the simulation results, however, in that the VHF cell patterns we observe maintain coherence over times of at least 2 – 3 hours (see Figures 10 and 14). It is clear that the VHF oscillatory power is intermittent (Figures 3d and 6), but it is likely that the bursts of power tend to recur at the same location. Visual inspection of Figures 2 and 6 lend at least psychological support to this idea.

In a recent review of solar convection Nordlund, Stein, and Asplund (2009) point out that convection near the solar surface is driven by the downflows of cool, dense plasma in the lanes and vertices around the granular patterns of rising, hot plasma. Because the pressure scale heights near 0.3 Mm at the top of the convection zone are smaller than the horizontal sizes of the convection cells, conservation of mass requires that there be strong outflows from the rising hot plasma into the sinking downflows around them. The larger the rising cell is horizontally, the faster the necessary outflow. Limiting the outflow to be slower than the 7 km s$^{-1}$ sound velocity sets a maximum horizontal diameter for the cells at about 4 Mm (Nelson and Musman, 1978; Nordlund, 1978). It is also the larger granular cells that develop cool centers that reverse the vertical flow and produce "exploding" (or "fragmenting") granules.

Further, it is pointed out that fluctuations in rising and expanding granular cells, that are becoming less dense, are damped out. Conversely, the cooler, sinking plasma in the intercellular downflows is compressed and increases in density, thus amplifying fluctuations and producing strong turbulence.

Putting these items together: (1) strongest outflows in the larger, exploding granules, (2) size scales from 2 – 4 Mm for these granules, and (3) compression of the intergranular downflows leading to turbulence, seems to account for the size scale of the cells outlined by our VHF G-band oscillatory power as well as the fact that intensity oscillations are preferentially found in the lanes and vertices around these cells. Finally, we re-emphasize that the cellular pattern of the oscillations is easiest to see in the VHF images, but the spectra of Figures 7, 8 and 9 show that more oscillatory power actually is present at these locations at lower frequencies.

**Acknowledgments**  *Hinode* is a Japanese mission developed and launched by ISAS/JAXA, collaborating with NAOJ as a domestic partner, NASA and STFC (UK) as international partners. Scientific operation of the Hinode mission is conducted by the Hinode science team organized at ISAS/JAXA. This team mainly consists of scientists from institutes in the partner countries. Support for the post-launch operation is provided by JAXA and NAOJ (Japan), STFC (U.K.), NASA, ESA, and NSC (Norway). The authors are grateful to Dr. Thomas E. Berger for helpful discussions of the *Hinode/SOT-FG* data. Wavelet software was provided by C. Torrence and G. Compo and is available at URL: http://paos.colorado.edu/research/wavelets/.